\documentclass[fleqn,10pt]{wlscirep}

\usepackage[utf8]{inputenc}
\usepackage[T1]{fontenc}
\usepackage{amsmath,amssymb,amsfonts,bm}
\usepackage{nameref,hyperref}
\usepackage{graphicx}
\usepackage{float}
\usepackage[aboveskip=1pt,font=footnotesize,labelfont=bf,labelsep=period,justification=raggedright,singlelinecheck=off]{caption}
\usepackage{comment}
\usepackage{siunitx}

\usepackage{subcaption}
\usepackage{chemfig}

\newcommand{\ca}{Ca$^{2+}$}  
\newcommand{\ip}{IP$_3$}
\newcommand{\ipr}{IP$_3$R}
\newcommand{\iprs}{IP$_3$Rs}

\title{Qualitatively Distinct Signaling in Cells: The Informational Landscape of Amplitude and Frequency Encoding. }

\author[1,2]{Alan Givr\'e}
\author[3,4,*]{Alejandro Colman-Lerner}
\author[1,2,*]{Silvina Ponce Dawson}
\affil[1]{Departamento de
  F\'\i sica, Facultad de Ciencias Exactas y Naturales, UBA, Buenos Aires, Argentina}
\affil[2]{Instituto de F\'\i sica de Buenos Aires (IFIBA), CONICET-UBA, Buenos
  Aires, Argentina}
\affil[3]{Department of Physiology, Molecular and Cellular Biology, School of Exact and Natural Sciences, University of Buenos Aires, Buenos Aires, Argentina}
\affil[4]{Institute of Physiology, Molecular Biology and Neurosciences, National Scientific and Technical Research Council (IFIBYNE-CONICET), Buenos Aires, Argentina}

\affil[*]{colman-lerner@fbmc.fcen.uba.ar; silvina@df.uba.ar}

\begin{abstract}
 Cells continuously sense their surroundings to detect modifications
  and generate responses. Very often changes in extracellular
  concentrations initiate {\it signaling cascades} that eventually
  result in changes in gene expression. Increasing stimulus strengths
  can be encoded in increasing concentration amplitudes or increasing
  activation frequencies of intermediaries of the pathway. In this
  paper we show that the different way in which {\it amplitude} and
  {\it frequency} encoding map environmental changes endow cells with
  qualitatively different information transmission capabilities.
  While amplitude encoding is optimal for a limited range of stimuli
  strengths, frequency encoding can transmit information with equal
  reliability over much broader ranges. The qualitative difference
  between the two strategies stems from the scale invariant
  discriminating power of the first transducing step in frequency
  codification.  The apparently redundant combination of both
  strategies in some cell types may then serve the purpose of
  expanding the span over which stimulus strengths can be reliably
  discriminated. In this paper we discuss a possible example of this
  mechanism in yeast.
\end{abstract}

\begin{document}
\flushbottom

\maketitle
\thispagestyle{empty}

\section*{Introduction}

Cells continuously sense their surroundings to detect modifications
and generate appropriate responses.  Environmental changes are often
due to changes in concentrations which induce further changes in
intracellular components producing a {\it signaling cascade}. There
are different strategies that cells use to ``interpret'' and react to
environmental changes.  On occasions, the intensity of the external
stimulus is encoded in the {\it amplitude} of the concentration of
active molecules in the following steps of the signaling
pathway~\cite{VenturaE3860,Poritz,amplitude5}. In others, it is
encoded in the {\it frequency} with which the molecules of one or more
steps switch between being active and
inactive\cite{de_koninck_science_1998,Albeck2013,Cai2008,Hao2012}.
When the end response involves modifications of gene expression these
different strategies may result in different dynamics of transcription
factor (TF) nuclear translocation. Namely, even upon stepwise changes
in the concentration of extracellular effectors, the nuclear TF
fraction can remain elevated for a certain time (mimicking the
dynamics of the environment) or display a non-trivial pulsatile
behavior~\cite{Yissachar2012,DALAL20142189, Cai2008,Hao2012}.  The
question then arises of how the two types of encoding differ and under
what circumstances one of them could be better suited than the
other~\cite{Cai2008,tostevin_2012,micali_pcbi_2015,dynamics_accurate_transmission_2014}.

The studies of~\cite{Hao2012, Hansen, informationoshea} showed that
cells may use the same TF to modulate the expression of different sets
of genes depending on the TF's nuclear translocation dynamics
(amplitude or frequency). Dynamics may then serve the purpose of {\it
  multiplexing} information transmission~\cite{multiplexing}.  What
matters in this case is whether different external stimuli can be
reliably encoded in different dynamics of TF's nuclear fractions
allowing their identification~\cite{informationoshea}.
In~\cite{givre_sci_rep_2023} we approached this problem applying
information theory to the steps that go from the TF's nuclear fraction
to mRNA production.  Using a two state promoter model we found that
the parameters that maximized information transmission lied in the
same bulk part for both amplitude and frequency encodings and that the
two strategies mainly differed in their sensitivity to changes in
promoter parameters~\cite{givre_sci_rep_2023} making frequency
modulation better suited for signal identification without the need to
incorporate extra regulatory motifs.  In the present paper we study
the differences and similarities between both encodings when the
mapping from the external stimulus to the TF's nuclear dynamics is
added to the model. The main result derived from this study is that
the different way in which amplitude and frequency encodings map
external stimuli results in qualitatively different information
transmission capabilities: while amplitude encoding is optimal for a
limited range of stimuli strengths, the information transmission
capability of frequency encoding can remain relatively invariant with
stimulus strength, provided that the differences between the resulting
frequencies are not filtered by the slower timescales of the
subsequent processing and that the observation time is long
enough. The combination of both mechanisms to encode the same type of
stimulus, which has been observed in certain systems, can then serve
to enlarge the range over which stimulus strengths can be reliably
discriminated. In this paper we discuss one possible such example in
{\it S. cerevisiae} cells.

Expanding signaling capabilities with dynamics~\cite{PURVIS2013945} is
characteristic of the universal second messenger calcium, \ca, which
encodes different inputs in different spatio-temporal distributions of
its free cytosolic concentration~\cite{Nelson1995, Nishiyama2000,
  berridge2000versatility} and differentially regulates gene
expression depending on its
dynamics~\cite{Dolmetsch1998,calcium_neuron_gene_pnas_2001}.
Interestingly, \ca\ is involved in various signaling pathways that
result in TF's nuclear fractions pulsatile behaviors~\cite{Cai2008,
  Yissachar2012,Carbo15022017}. Although TF oscillations might not be
a mere reflection of those of intracellular \ca, they share some
common properties. Sequences of intracellular \ca\ pulses elicited by
constant concentrations of external effectors have been observed to be
very stochastic in different cell types, particularly in those in
which \ca\ release from the endoplasmic reticulum through Inositol
1,4,5-trisphosphate (\ip) receptors (\iprs) is
involved~\cite{Skupin2008,dragoni_2011,Thurley2014}.  It was
argued~\cite{Thurley2011} that this stochasticity occurs because
pulses arise via random \ca-channel openings which yield localized
\ca\ elevations that eventually nucleate to produce a global increase
in \ca. This behavior is characteristic of spatially extended
excitable systems in which ``extreme events'' (\ca\ spikes or pulses)
are triggered by noise and subsequently amplified through
space~\cite{Lopez2012,PhysRevResearch.3.023133}.  We briefly remind
here that excitable systems are characterized by a stable stationary
state and a threshold which, if surpassed due to a perturbation, a
long excursion in phase space (a spike) is elicited before the system
relaxes to its stable fixed point~\cite{izhikevich}. Excitability has
been associated with intracellular
\ca\ patterns~\cite{lechleiter_science,lechleiter_cell_1992,lirinzel,
  Tang7869} and with the dynamics of some pulsatile TFs, {\it e.g.},
the tumor supressor,
p53~\cite{p53_excitable,p53_excitable_network_falcke}. Models of the
way in which the TF, Msn2, responds to different stresses in yeast
demonstrated that bursts of Msn2 nuclear localization arise from noise
in the signaling pathways~\cite{msn22} and noise-triggered excitable
systems have been found to describe correctly differentiation in
bacteria~\cite{functional_role_noise}.  In the case of \ca\ pulses,
the interspike time intervals have been observed to be the sum of a
fixed component (due to spike duration and refractoriness) and a
stochastic one of average, $T_{av}$, that decreases exponentially with
the effector's concentration~\cite{Thurley2014}. We have recently
derived this dependence~\cite{Givre2024} using a simple model for the
dynamics of intracellular \ca, in the excitable regime and subject to
noise. Using experimental observations~\cite{Marchant2001} to estimate
the noise amplitude we derived the exponential dependence both through
the use of Kramer's law for thermally activated barrier
crossings~\cite{kramers} and via numerical simulations. Kramer's law
prescribes an exponential dependence of the waiting time for barrier
crossing with the height of the barrier and the noise
intensity. In~\cite{Givre2024} we showed how, in the case of
\ipr-mediated \ca\ pulses, increasing the external stimulus strength
reduces the barrier height and increases the noise, resulting in a
mean interpulse frequency which increases exponentially with the
stimulus.  As discussed in~\cite{Givre2024}, a similar result can be
obtained for other noisy excitable
systems~\cite{Eguiamindlin2000}. These studies lead us to the
conclusion that, if noise and excitability are involved, we can expect to find an
exponential dependence between the mean interpulse frequency and the
intensity of what can be associated with a stimulus.  Interestingly,
the TF, Crz1, in yeast exhibits bursts of nuclear
localization~\cite{Cai2008} whose mean frequency can be shown to
increase exponentially with extracellular \ca, the external effector
in this case. Other TFs that exhibit pulsatile nuclear localization
have frequencies that are convex increasing functions of the external
stimulus strength~\cite{DALAL20142189,Hao2012} and might, in
principle, depend exponentially on such strength.  Alhtough we have
not seen a thourough analysis of this dependence outside the realm of
\ca\ signaling, based on this discussion, here we
assume that frequency encoding entails an exponential dependence of
mean frequency with external input
strength~\cite{givre_2018,Givre2024}.

In this paper we use a simple two-step model to compare the
information transmission capabilities of amplitude and frequency
encoding upon a constant external stimulus. The first step of the
model goes from the stimulus to the TF's nuclear fraction with a
different mapping and nuclear TF dynamics depending on the
codification. The second step models transcription and goes from the
nuclear TF concentration to the mRNA produced over a fixed time frame.
The only difference between amplitude and frequency encoding in this
step is due to the different nuclear TF dynamics. For the
transcription step we use a two-state promoter model with a TF
dependence of the transition rates originally obtained from data
fitting~\cite{Hansen} which we derived from a mechanistic
model~\cite{givre_sci_rep_2023} and parameter values that are based on
our previous studies~\cite{givre_sci_rep_2023}. We then compute the
mutual information, $MI$, between the mRNA produced and the stimulus
strength, $I_{ext}$, using various $I_{ext}$ distributions that are
defined over the same compact support but differ in their mode, mean
and median while keeping approximately the same variance. The first
indication of the qualitative difference between amplitude and frequency
encoding is reflected in the different dependence of $MI$
 with the median of the $I_{ext}$ distribution obtained numerically.  We then
derive analytic results which
show that the reason for this difference can be traced back to the
different properties of the mapping from stimulus strength to
nuclear TF's fraction of both encodings which endows frequency
codification with a scale invariant discriminating power. We then study how the subsequent steps in the
processing of the response can limit this scale invariance identifying two key limiting timescales.  We finally discuss how the combination of both
encodings can expand the range over which external stimulus strengths
can be reliably discriminated focusing on a possible example in the
pheromone respose pathway in cells of {\it S. cerevisiae}. 

\section*{Methods}
\label{sec:methods}

\subsection*{The model.}

\begin{figure}[ht!]
  \centering
\includegraphics[width=520pt]{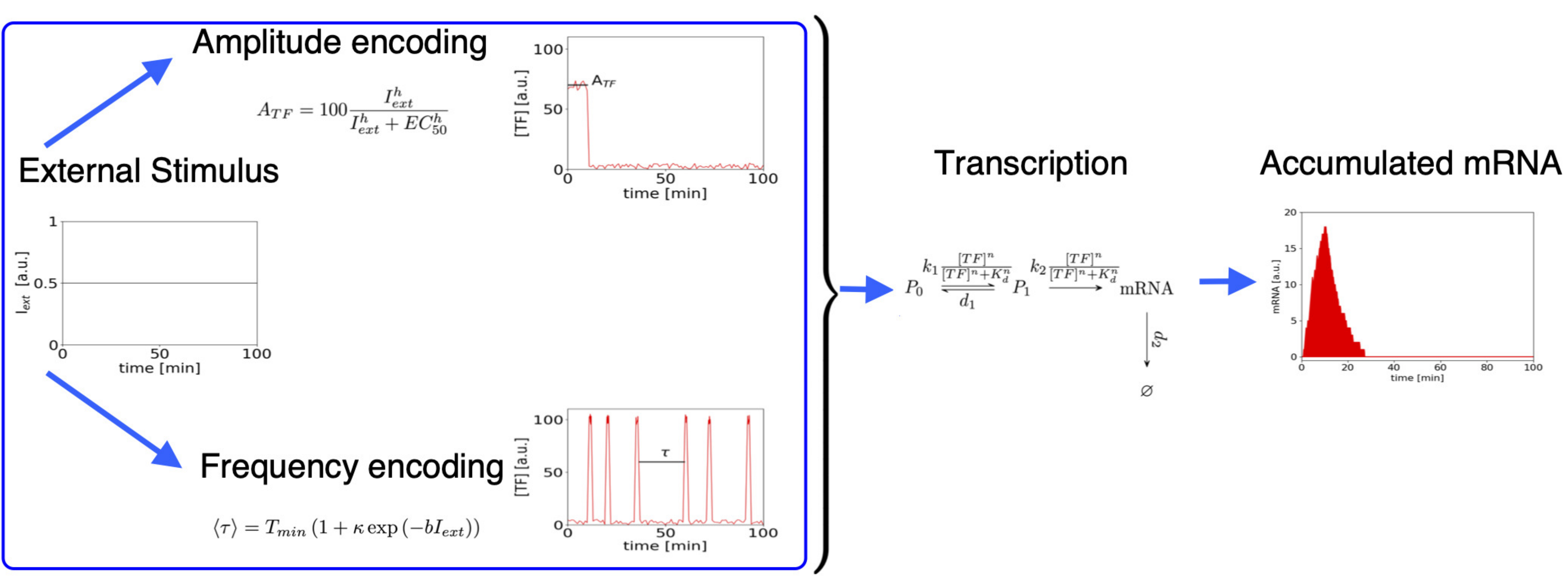}
    \caption{The model. The external stimulus strength, $I_{ext}$,
      determines the mean concentration, $A_{TF}$, of the single TF pulse and the
      mean time, $\langle \tau\rangle$, between successive pulses in amplitude and frequency encoding, respectively.   The nuclear TF concentration, $[TF]$, modulates the transcription rate according to the two-state model depicted  which, in turn, determines the mRNA time integral that is taken as the indicator of gene expression. The transcription model is the same as the one considered in~\cite{givre_sci_rep_2023}.
 }
\label{fig:model}
\end{figure}

We show in Fig.~\ref{fig:model} a scheme of the model considered in the paper.
A constant external stimulus can either elicit a single pulse of TF nuclear traslocation (amplitude encoding) or a sequence of pulses (frequency encoding). 
In the former, the TF nuclear concentration ($[TF]$) pulse has
mean amplitude, $A_{TF}$, and in frequency encoding the pulses are separated by a stochastic interpulse time, $\tau$. The external stimulus strength, $I_{ext}$,
determines $A_{TF}$ and $\langle \tau\rangle$ for each encoding. We have not included  a detailed dynamical system to derive TF from $I_{ext}$ since it would only add to the transient of the TF dynamics without affecting the mRNA time integral that we take as the output of the process. $[TF]$ modulates the transcription rate according to the two-state model included in Fig.~\ref{fig:model}~\cite{Hansen,givre_sci_rep_2023}, where $P_0$ and $P_1$ represent the two (effective) states of the promoter and transcription proceeds only when the promoter is in the $P_1$ state at rate, $k_2 [TF]^nP_1/(K_d^n+ [TF]^n)$ (as explained later, we also use $P_0$ and $P_1$ to denote the probabilities that the system is in one or the other state), while mRNA is degraded at rate, $d_2$. This part of the model as well as the indicator of gene expression (the accumulated mRNA)
are the same as the ones we used in our previous study~\cite{givre_sci_rep_2023} where we performed an extensive search of optimal parameter values. The parameters of the transcription step used in the present paper were chosen based on this previous study.  We describe the model in more detail in what follows.



\subsubsection*{Transcription.}
The transcription step common to frequency and amplitude encoding is given by~\cite{Hansen,givre_sci_rep_2023}:
\begin{align}
   & \dot{P}_1=\frac{k_1 [TF]^n}{K_d^n+[TF]^n}-\left(\frac{k_1 [TF]^n}{K_d^n+[TF]^n}+d_1\right)P_1,\label{eq:P_0}\\
     &X=N\xrightarrow[]{\frac{{k_2} [TF]^nP_1}{K_d^n+ [TF]^n}}{X=N+1},\,
    {X=N}\xrightarrow[]{d_2 X}X=N-1
\label{eq:mRNA} ,
\end{align}
where $[TF]$ is the nuclear TF concentration (in dimensionless units),
$P_1$ represents the probability that the promoter is in its active
TF-bound state and $X(t)$ is the random number of mRNA molecules at
time, $t$, that can take on the values $N=\{0,1,2\cdots\}$. 
Eq.~(\ref{eq:P_0}) is deterministic and Eq.~(\ref{eq:mRNA}) is
stochastic. In the simulations we solve Eq.~(\ref{eq:P_0}) using
Euler's method with time step $dt=0.1min$, integration time $T=100min$, and $n=10$, $K_d=40$ (in the
same dimensionless units as $[TF]$), $k_1=1/min$, $d_1=0.01/min$,
$k_2=10/min$ and $d_2=0.12/min$.  In all cases, the output is:
\begin{align}
       O= \int_0^T dt\,  X(t)
\label{eq:Out} .
\end{align}

\subsubsection*{From the external stimulus to the nuclear TF.}
For amplitude
modulation, $[TF]$ is modeled as a single pulse of 10 min duration
and (dimensionless) amplitude:
\begin{equation}
   A=A_{TF}(I_{ext})+\zeta,
\label{A_random}
\end{equation} 
 with $\zeta$ a Gaussian distributed
random variable 
of standard
  deviation, $\sigma_\zeta=10$,
 and
$A_{TF}$ related to the external input strength, $I_{ext}$, (typically, the dimensionless concentration
of an external ligand),
 by a cooperative Hill function:
\begin{equation}
    A_{TF}(I_{ext})=\langle A\rangle=100\frac{I_{ext}^{h}}{I_{ext}^{h}+EC_{50}^{h}} ,
\label{ATF}
\end{equation}
which aggregates in one step the various processes that go from the
external stimulus to the TF's nuclear fraction.  $EC_{50}$ and $I_{ext}$ are measured
in the same dimensionless units, which not necessarily coincide with those of $[TF]$.

For frequency
modulation, $[TF]$ is modeled as a sequence of 1 min-duration
square pulses of (dimensionless) amplitude 100 plus a random variable, $\zeta$, as
in the case of amplitude encoding, and stochastic
interpulse time intervals, 
\begin{equation}
   \tau= T_{min}+\eta,
\label{tau_random}
\end{equation} 
with $T_{min}$ fixed~\cite{Thurley2014} and $\eta$ 
exponentially distributed with rate parameter
exponentially dependent on $I_{ext}$~\cite{givre_2018,Givre2024} so that:
\begin{align}
  T_{IP}(I_{ext})&\equiv\langle \tau\rangle =T_{min}\left(1+\kappa\exp\left(-bI_{ext}\right)\right),
\label{FTF}
\end{align}
with $\kappa, b >0$.  In this case $b$ is measured in the inverse of the (dimensionless) units of $I_{ext}$. 

\subsection*{Input, Output and Mutual Information.}

\begin{figure}[ht!]
  \centering
\includegraphics[width=500pt]{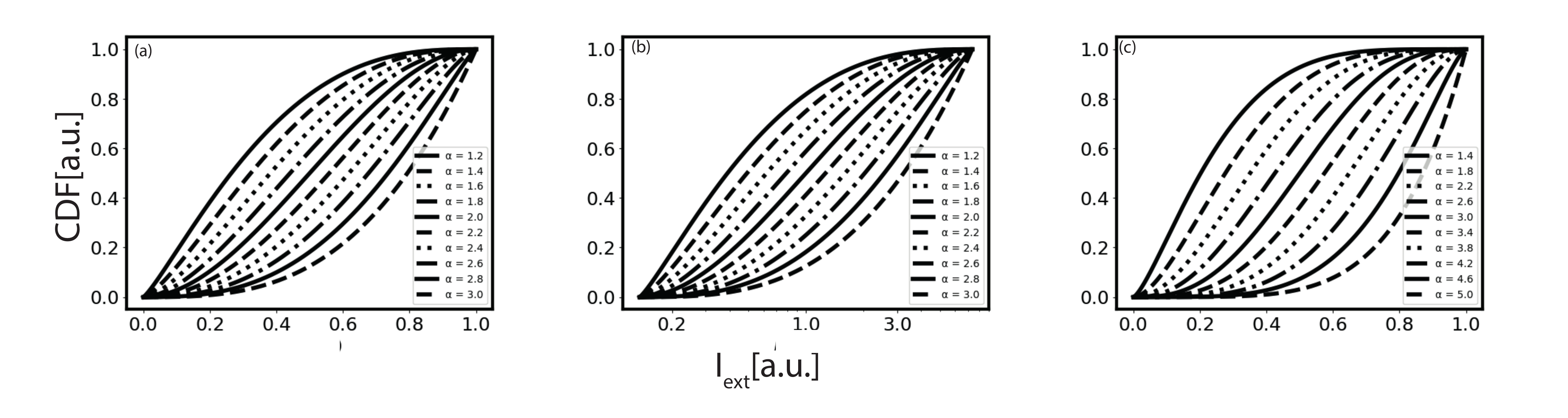}
    \caption{Cumulative distribution function (CDF) of the 
      stimulus strength, $I_{ext}$, obtained with Eq.~(\ref{lin}) ((a), (c)) and Eq.~(\ref{log}) (b) combined with
      Eq.~(\ref{f_x}) for 10 $(\alpha, \beta)$ pairs  such
      that $\alpha+\beta =4$ ((a), (b)) and $\alpha+\beta =6$ ((c)). Varying $\alpha$ and $\beta$ for increasing values of
     $\alpha+\beta\ge 3$ yields broader  ranges of the $I_{ext}$ median and standard 
deviation (with smaller values for the latter: $\sim 0.19-0.22$ in (a) and $\sim 0.14-0.19$ in (b)). Values such that $\alpha+\beta = 4,5$ present a good balance of medians that cover relatively well the [0,1] interval with relatively invariant and not too small variances.  }
\label{fig:I_ext_vs_dist}
\end{figure}

Using an $I_{ext}$ distribution of compact
support, $[I_m,I_M]$, we compute numerically the mutual information,
MI, between $I_{ext}$ and the mRNA time integral, $O$, given by Eq.~(\ref{eq:Out}) over the $T=100 min$
duration of the simulation using the
Jackknife method~\cite{jackknife1,jackknife2}.  We have used $T_{min}=5, 10\, min$, various
values of $\kappa$ and, for the $I_{ext}$ distribution either one of these expressions:
\begin{align}
  I_{ext} &= x,\label{lin}\\
  I_{ext} &= \exp\left(4(x-0.5)\right)\label{log}
  \end{align}
 with $0\le x\le 1$ Beta-distributed:
\begin{align}
  f(x) = \frac{x^{\alpha-1}(1-x)^{\beta-1}} {\mathrm {B} (\alpha,\beta)},
\label{f_x}
\end{align}
and different choices of $\alpha, \beta>1$ s.t.  $\alpha + \beta \ge 3$,
so as to obtain different values for
the median of $I_{ext}$. 
Choosing Eqs.~(\ref{lin}) or (\ref{log})  the corresponding
range of $I_{ext}$ values is spanned ``linearly'' or ``logarithmically'',
respectively, with relatively uniform ``linear'' or ``logarithmic''
standard deviation as a function of the median. Within those ranges,
increasing values of $\alpha+\beta\ge 3$ yield more widely spread
medians and  smaller values of the standard deviation as
illustrated in Fig.~\ref{fig:I_ext_vs_dist}. In the paper we show the results obtained
with the distributions of Fig.~\ref{fig:I_ext_vs_dist}~(a) because their medians cover
reasonably well the whole interval, $[0,1]$, with a relatively invariant and not too small standard 
deviation (between 0.19 and 0.22). 


\subsection*{Computation of the distribution function of the intermediaries of the response.}

In the approach followed in this paper the intermediaries of the response are the amplitude, $A$,  or the interpulse time, $\tau$, (or, equivalently, the frequency $\nu\equiv1/\tau$) of the TF's nuclear concentration in the case of amplitude and frequency encoding, respectively.  The amplitude, $A$, is given by Eqs.~(\ref{A_random})--(\ref{ATF}) where $\zeta$ is normally distributed with standard deviation, $\sigma_\zeta$. This implies that $A$ is Gaussian distributed with mean $A_{TF}$ and standard deviation, $\sigma_\zeta$. 
The interpulse time, $\tau$, is given by Eq.~(\ref{tau_random}) with $\eta$ exponentially distributed with mean, $\langle \eta \rangle = T_{IP}-T_{min}$, where $T_{IP}$ is given by Eq.~(\ref{FTF}). $A$ and $\tau$ (or $\nu$) can be thought of as the ``input'' of the transcription model. Given that their distributions are known for a given value of the external stimulus, $I_{ext}$,  and that the $I_{ext}$ distribution is known as well, it is possible to derive the  $A$  and $\tau$ or $\nu$ distributions that are fed into the transcription model in our approach. In particular, in the paper we show the results obtained for the choice of $I_{ext}$ distribution given by Eqs.~(\ref{lin}) and (\ref{f_x}). In such a case, the cumulative distribution functions (CDF) of $A$ and $\nu$ can be computed as:
\begin{align}
CDF_{A}(A) = \int^A_0 da\int_0^1 dI_{ext} f(I_{ext}) \frac{e^{-\frac{\left(a-A_{TF}(I_{ext})\right)^2}{2\sigma_\zeta^2}}}{\sqrt{2\pi\sigma_\zeta}},\label{eq:CDFA}
\end{align}
\begin{align}
  CDF_{\nu}(\nu) = 1-\int^{1/\nu}_{T_{min}} dt \int_0^1 dI_{ext} f(I_{ext}) \frac{ e^{-\frac{t-T_{min}}{T_{IP}(I_{ext})-T_{min}}}}{T_{IP}(I_{ext})-T_{min}},\label{eq:CDFnu}
\end{align}
with $f$ given by Eq.~(\ref{f_x}), $A_{TF}$ by Eq.~(\ref{ATF}), $T_{IP}$ by Eq.~(\ref{FTF}) and where we are assuming in Eq.~(\ref{eq:CDFA}) that the integral in $a$ from $-\infty$ to 0 is negligible for most values of $A$. We use these
expressions to compute numerically these two CDFs. 

\section*{Results}

\subsection*{Amplitude and frequency encodings yield qualitatively different dependences between MI and stimulus strength.}
\label{subsec:qualitative}

\begin{figure}[ht!]
  \centering
\includegraphics[width=250pt]{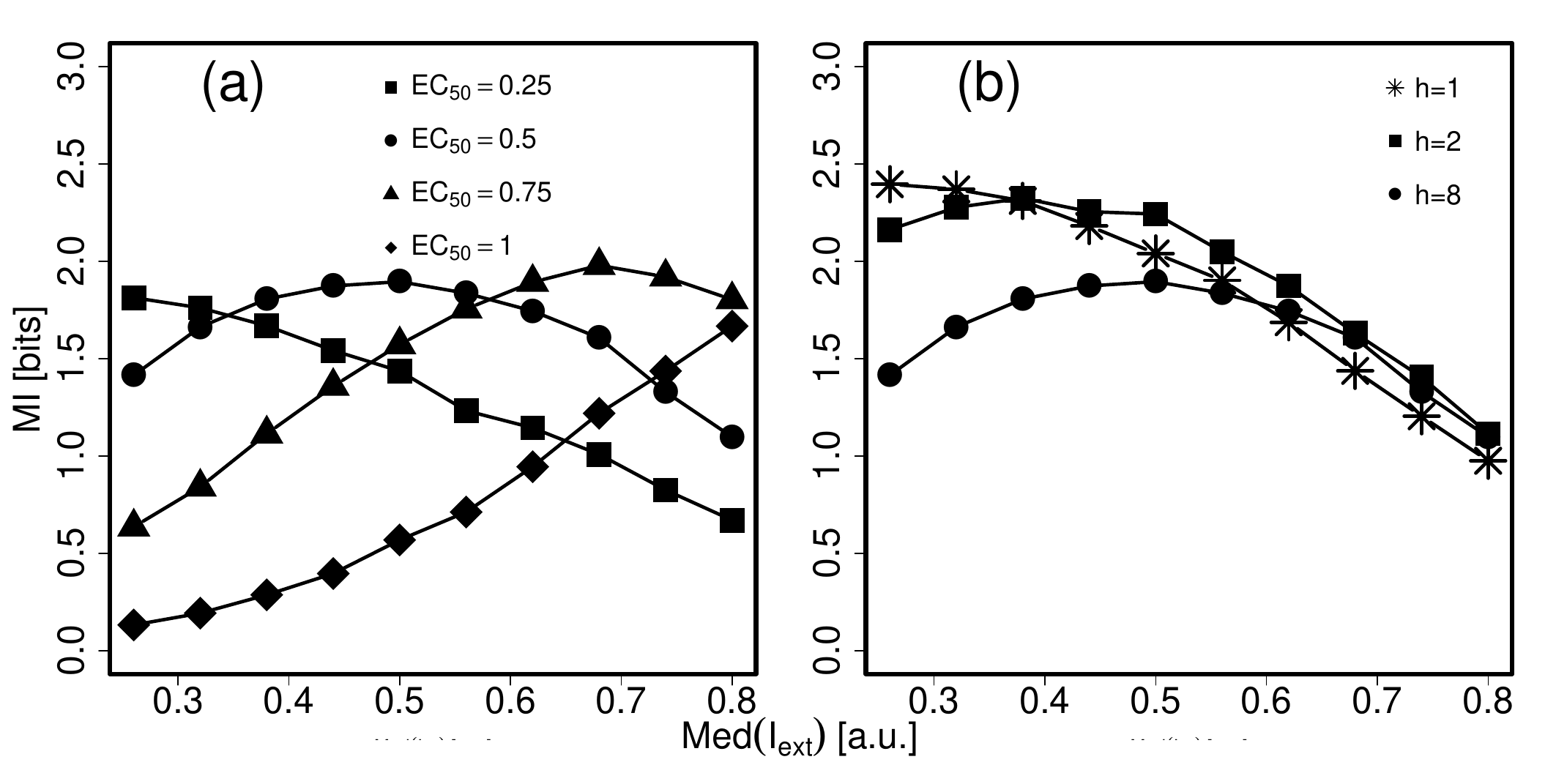}
    \caption{Mutual Information between $O$ (Eq.~(\ref{eq:Out})) and
      $I_{ext}$, for amplitude encoding and the 10 $I_{ext}$ distributions of
      Fig.~\ref{fig:I_ext_vs_dist}~(a), as a function of Med($I_{ext}$).  Eq.~(\ref{ATF}) was used with $h=8$
      and $EC_{50}=0.25, 0.5, 0.75, 1$ in (a) and with $EC_{50}=0.5$
      and $h= 1, 2, 8$ in (b).  The results are depicted with symbols and joined by curves for the sake of clarity.  }
\label{fig:MI_amp}
\end{figure}

We show in Figs.~\ref{fig:MI_amp} the values, MI,
obtained for amplitude encoding using the 10 $I_{ext}$ distributions
of Fig.~\ref{fig:I_ext_vs_dist}~(a), plotting MI as a function of the
corresponding medians, Med$(I_{ext})$, for different choices of $h$ and $EC_{50}$ in
Eq.~(\ref{ATF}). Qualitatively similar results were obtained for the
distributions of Fig.~\ref{fig:I_ext_vs_dist}~(b) and for other
parameter values. We
observe in Fig.~\ref{fig:MI_amp} that MI is maximum at Med$(I_{ext})\le EC_{50}$ which approaches $EC_{50}$ from below as $h$
increases (Fig.~\ref{fig:MI_amp}~(b)). It is also apparent that MI remains within a small percent of this maximum for a
limited range of Med$(I_{ext})$ (around $EC_{50}$ for $h$ large enough, as illustrated in Fig.~(\ref{fig:MI_amp}~(a)).

\begin{figure}[ht!]
  \centering
\includegraphics[width=250pt]{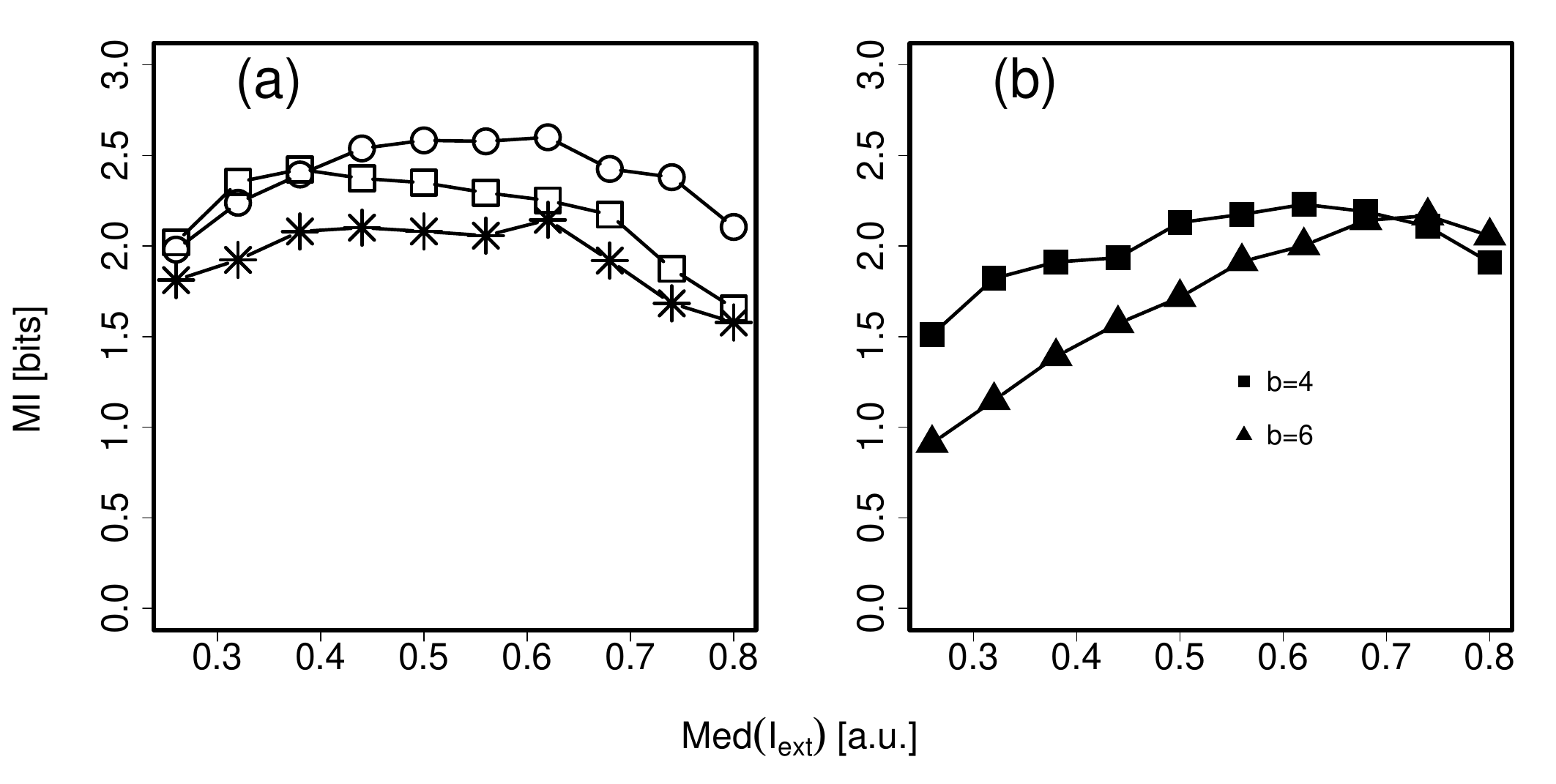}
    \caption{Similar to Fig.~\ref{fig:MI_amp}, but for frequency
      encoding. In this case,  Eq.~(\ref{FTF}) was used  with
      $T_{min}=5min$, $b=4$, $\kappa = 9$ (squares), $T_{min}=5min$, $b=6$, $\kappa = e^4$ (circles) and
      $T_{min}=10min$, $b=4$ and $\kappa = 9$ (asterisks) in (a) and with
      $T_{min}=5min$, $b = 4, 6$ and $\kappa = \exp(b)$ in (b).  }
\label{fig:MI_freq}
\end{figure}

The situation observed in Fig.~\ref{fig:MI_amp} is qualitatively
different from the one derived for frequency encoding provided that
$\kappa$ and $b$ in Eq.~(\ref{FTF}) are such that the bulk of the
$T_{IP}$ distribution allows the discrimination of nearby mean
frequencies and includes values that are not too large so that the
probability of eliciting one pulse during the finite time of the
simulation ($100min$) is non-negligible. These
two conditions can be satisfied simultaneously, as illustrated in
Fig.~\ref{fig:MI_freq}~(a) where we observe that MI can remain close
to its maximum value for a broader range of external input strengths
than in the case of amplitude encoding.  We will discuss later under
what conditions MI for frequency encoding can remain relatively
constant as Med$(I_{ext})$ is varied, something that is not always satisfied
as illustrated in Fig.~\ref{fig:MI_freq}~(b). We focus now on the reasons that
underlie the qualitative differences between amplitude and frequency
encoding which limit the information transmission capabilities of the former
to a narrower range of stimulus strengths than for the latter.

\subsection*{The qualitative difference between 
amplitude and frequency encoding can be attributed to the different way in which
they map the environment.}
\label{subsec:expla}

The different dependence of MI with Med$(I_{ext})$ for
amplitude and frequency encoding can be traced back to the different way in which
the set of external inputs is ``mapped'' on the subsequent steps.
As we show in what follows, while a mapping in the form of Eqs.~(\ref{A_random})--(\ref{ATF}) only
allows to discern a relatively narrow set of external inputs around or below
$EC_{50}$, Eqs.~(\ref{tau_random})--(\ref{FTF}) can map the whole set of external inputs onto a
set of discernible interpulse time distributions. 

Let us consider two
nearby values, $I_{ext}$ and $I_{ext}+\Delta_I$. In the case of amplitude encoding (Eqs.~(\ref{A_random})--(\ref{ATF})), these two values
will yield two Gaussian distributions of standard deviation, $\sigma_\zeta$, centered around $A_{TF}(I_{ext})$ and $A_{TF}(I_{ext}+\Delta_I)$, respectively. This first step  will allow  $I_{ext}$ and $I_{ext}+\Delta_I$  to be distinguishable if the ratio,
\begin{align}
  \Delta_A&\equiv \frac{A_{TF}(I_{ext}+\Delta_I)-A_{TF}(I_{ext})}{2\sigma_\zeta} , \label{ratio_A}
\end{align}
 satisfies
\begin{align}
  \Delta_A&\approx \frac{\Delta_I}{EC_{50}}\, \frac{100 h}{2\sigma_\zeta}\frac{I_{ext}^{h-1}/EC_{50}^{h-1}}{(I_{ext}^{h}/EC_{50}^{h}+1)^2}>1 . \label{r_A}
\end{align}
In the case of frequency
encoding (Eqs.~(\ref{tau_random})--(\ref{FTF})) $I_{ext}$ and $I_{ext}+\Delta_I$ will yield
two exponential distributions for the ``shifted'' interpulse time, $\tau-T_{min}$ of mean
$T_{IP}(I_{ext})-T_{min}$ and $T_{IP}(I_{ext}+\Delta_I)-T_{min}$, respectively.  These two distributions  will be distinguishable if the quantiles, $p$ and $1-p$,  of each of them
be further apart. This is satisfied if
\begin{align}
 \Delta_F&\equiv \exp(b\Delta_I)\frac{\log(p)}{\log(1-p)} > 1,      \label{r_F}
\end{align}
for some $p>1/2$ (e.g., $p=3/4$ guarantees that the overlap of the two
distributions does not exceed 1/4 of the total probability).

Eqs.~(\ref{r_A}) and (\ref{r_F}) are qualitatively different:
$\Delta_A$ is a non-monotone function of $I_{ext}$, while $\Delta_F$
does not depend on $I_{ext}$. For $h=8$, for example, the term
multiplying $ {\Delta_I}/{EC_{50}}$ in Eq.~(\ref{r_A}) attains its
maximum value at $x\equiv I_{ext}/EC_{50}\approx 1$ and decays by 50\%
at $x\approx 0.77$ and $x\approx 1.20$. Thus, the first step of
amplitude encoding (Eqs.~(\ref{A_random})-(\ref{ATF})) for $h=8$ would distinguish values,
$I_{ext}$, that differ among themselves by $\sim 0.2 EC_{50}$ if
$0.77<I_{ext}/EC_{50}<1.20$. This is consistent with the behavior of
MI $vs$ Med$(I_{ext})$ in Fig.~\ref{fig:MI_amp}~(a).
A similar discernment is achieved over
the ranges $I_{ext}/EC_{50}\le 0.4$ for $h=1$ and $0.17<I_{ext}/EC_{50}<1.4$ for $h=2$, consistently with the results of Fig.~\ref{fig:MI_amp}~(b). 
 Eq.~(\ref{r_F}), on the other hand, shows that for large
enough $b$, the first  step of frequency encoding, Eqs.~(\ref{tau_random})--(\ref{FTF}), will allow the discernment of nearby input
strengths with the same resolution across the whole range of $I_{ext}$
values. This qualitative difference is also found in the  Kullback–Leibler (KL) divergence (a measure of the statistical distance) between the  conditional distributions for $I_{ext}$ and $I_{ext}+dI$. Namely, KL
depends on $I_{ext}$ for amplitude encoding (KL=$2\Delta_A^2$ in this case)
while it does not for frequency encoding for which it reads:
\begin{align}
{\rm KL} &=  b\Delta_I+ \exp\left(b\Delta_I\right).\label{eq:KL}
\end{align}

\begin{figure}[ht!]
  \centering
\includegraphics[width=250pt]{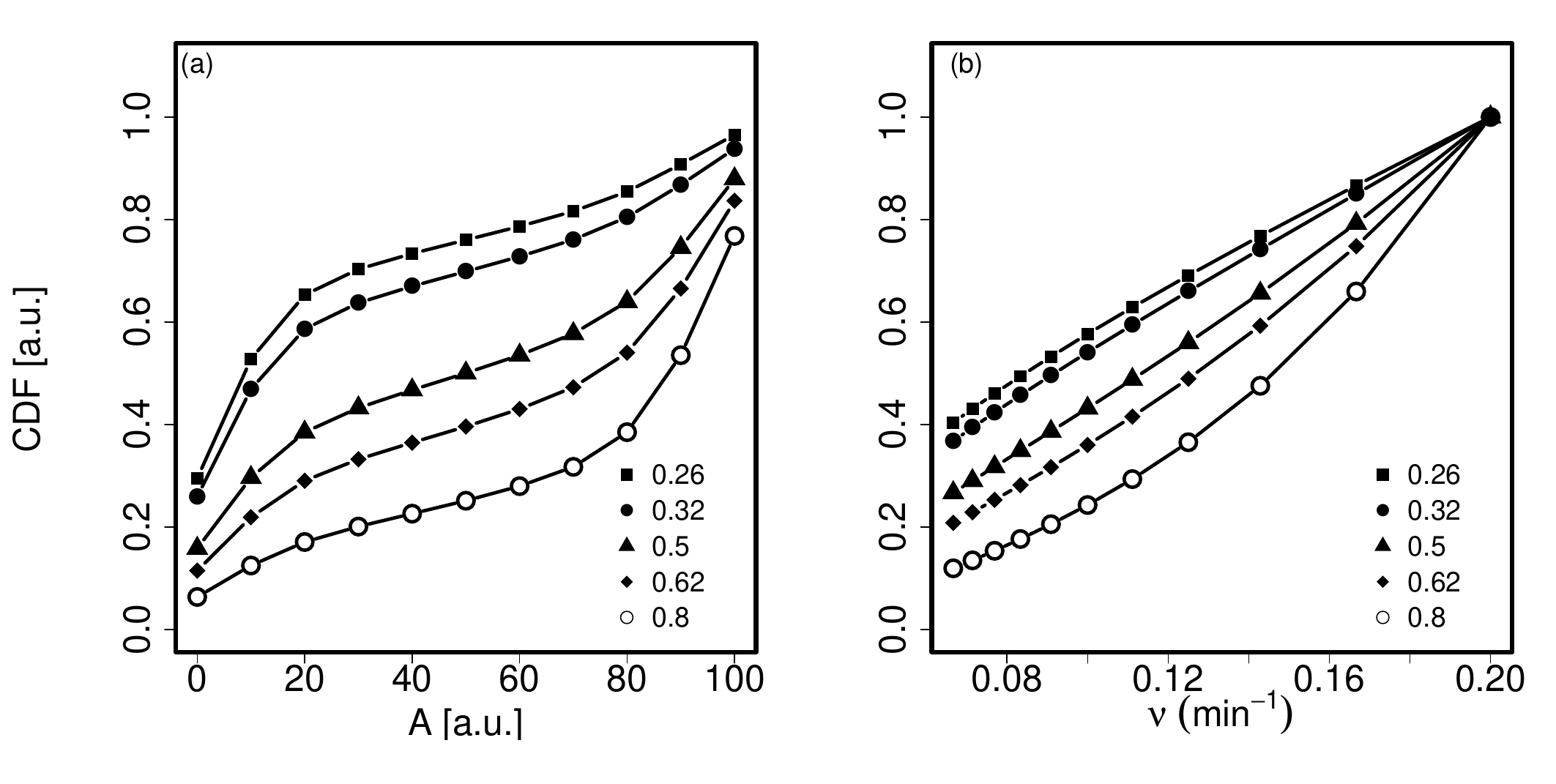}
    \caption{CDFs of the TF
      amplitude, $A$, (a) and of the interpulse frequency, $\nu=1/\tau$, (b)   derived
for 5 of the $I_{ext}$ distributions depicted in Fig.~\ref{fig:I_ext_vs_dist}~(a) using $EC_{50}=0.5$ and $h=8$ in (a) (same parameters as the curve with solid circles in Fig.~\ref{fig:MI_amp}~(a))
and $T_{min}=5min$, $b=4$ and $\kappa =9$ in (b) (same parameters as the curve with open squares in Fig.~\ref{fig:MI_freq}~(a)). The legends indicate the values of the corresponding  $I_{ext}$ medians.   }
\label{fig:CDFs}
\end{figure}

The different behavior just described is apparent in
Fig.~\ref{fig:CDFs} where we have plotted the CDFs of TF amplitude,
$A$, (a) and interpulse frequency, $\nu=1/\tau$, (b) computed for each
encoding type as explained in Methods with a given parameter set of
the $I_{ext}$-TF mapping ($h=8$ and $EC_{50}=0.5$ in (a) and $T_{min}=5min$, $b=4$ and $\kappa =9$ in (b))
and 5 $I_{ext}$ distributions whose medians
are indicated in the legend. Analyzing Fig.~\ref{fig:CDFs}~(a) in
terms of the MI that is eventually conveyed for amplitude encoding with
$h=8$ and $EC_{50}=0.5$  (curve with solid circles
in Fig.~\ref{fig:MI_amp}~(a)) we conclude that the $I_{ext}$ distribution that yields maximum MI for these parameters (the one
with Med$I_{ext}=$ 0.5) corresponds to the amplitude CDF which is
closest to that of a uniform distribution (solid triangles in
Fig.~\ref{fig:CDFs}~(a)). In the case of frequency encoding, the analogous
comparison should be made with the curve depicted with open squares in Fig.~\ref{fig:MI_freq}~(a) (which was obtained using $T_{min}=5min$, $b=4$ and $\kappa =9$). In this case,  almost
all the CDFs depicted in Fig.~\ref{fig:CDFs}~(b) are similarly close
to that of a uniform distribution over a certain support. This could
explain the weak dependence of MI with Med$(I_{ext})$ for the
curve depicted with open squares in Fig.~\ref{fig:MI_freq}~(a). The fact that the distribution of the intermediary of the response which yields maximum MI is almost uniform  ressembles the optimal input/output relation derived for
cases with small, independent of the mean,
noise~\cite{Laughlin_1981,Bialek_2018}.  This description, however, corresponds to the first step
in the generation of the response and other uncertainties are
subsequently added which further degrade the information. In particular, this is very relevant in the case of frequency encoding as we explain in the following Section.

\subsection*{The invariance of MI with stimulus strength in frequency encoding is limited by two key timescales.}
\label{subsec:times}

The examples of Fig.~\ref{fig:MI_freq} illustrate that MI for
frequency encoding can remain relatively invariant as Med($I_{ext}$)
varies but that it can also decrease for small or large values of the
median. This different behavior depends on some of the timescales of
the transcription step.  On one hand,  the finite
time of the simulations imposes a limit on the minimum frequency that
will likely lead to mRNA production. This limitation is also relevant
in physiological situations, due to the finite turn over time of
proteins and the need to generate responses  within a 
time frame.  In fact, MI decreases with Med($I_{ext}$) if the
probability of eliciting at least one pulse during the finite
observation time, $T=100min$ in our simulations, becomes too
small. This is particularly noticeable in the examples depicted in
Fig.~\ref{fig:MI_freq}~(b). A simple calculation done as if the
$I_{ext}$ distribution were concentrated around $I_m=$Med$(I_{ext})$
gives that the probability of eliciting at least one pulse, $\approx
1-\exp\left(-e^{bI_M}(T-T_{min})/(\kappa T_{min})\right)$, at
$I_m=0.27$ ($\alpha =1.2$ in Fig.~\ref{fig:I_ext_vs_dist}~(a)) is $
\sim 0.21$ and $\sim 0.64$ for the cases depicted with circles and
squares, respectively, in Fig.~\ref{fig:MI_freq}~(b) and $\sim 0.83$
and $\sim 1$ for those depicted with circles and squares,
respectively, in Fig.~\ref{fig:MI_freq}~(a).
On the other hand, too large input strengths can become
indistinguishable for the mRNA production if the difference between
their corresponding mean interpulse times, $T_{IP}$ (Eq.~(\ref{FTF})),
is so small that is filtered out by some of the slower processes of
the transcription step. Let us consider the examples depicted with
open symbols in Fig.~\ref{fig:MI_freq}~(a), both of which yield a
maximum MI$\sim 2.5$bits which occurs when using the distribution with
Med$(I_{ext})=0.44$ in one case (squares) and with
Med$(I_{ext})=0.5-0.62$ in the other (circles). The maximum $\sim
2.5$bits roughly means that $\sim 2^{2.5}$ intervals of $I_{ext}$ values can
be discerned for these examples when using their ``optimal''
distributions.  Let us now consider two external inputs, $I_{ext}$ and
$I_{ext}+\Delta_I$, that differ by the minimum discernible stimulus
strength under these optimal conditions, $\Delta_I =1/ 2^{2.5}\approx
0.18$. Given Eq.~(\ref{FTF}), these two inputs will produce interpulse
time distributions with means that differ by $\Delta T_{IP}\approx
\kappa T_{min}\exp(-b I_{ext})b\Delta_I$.  Replacing $I_{ext}$ by the
value, Med$(I_{ext})$, at which each of these examples yields maximum
MI, we estimate that the minimum discernible $\Delta T_{IP}$ for both
of them is $\sim 7-8 min$. A similar result is obtained for the
example depicted with asterisks in Fig.~\ref{fig:MI_freq}~(a) for
which the maximum $MI$ is $\sim $2 and occurs at Med$(I_{ext})\sim
0.62$.  The result obtained is approximately equal to the
characteristic mRNA degradation time of the simulations ($\sim 8 min$)
which, in turn, agrees with the fastest mRNA turnover times determined
in yeast~\cite{pnas_decay_mrna}. We have previously observed that this
timescale is key in limiting the information transmitted through the
transcription step (see Fig. 3D in~\cite{givre_sci_rep_2023}).  The
observation of indistinguishable mean interpulse times in experiments
(see {\it e.g.}, Fig. 4B in~\cite{Thurley2014} or Fig.~3C
in~\cite{Carbo15022017}) can then be used to determine the range of inputs
for which frequency encoding can work.

\subsection*{ 
Combining amplitude and frequency encoding to expand the range of distinguishable stimuli.
}
\label{subsec:whatfor}

The different way in which the two types of strategies encode external
stimuli might serve to enlarge the range of distinguishable stimulus
strengths in cell types that use the two encodings to respond to the
same type of stimulus. This could happen in the yeast mating
response in which the two types of encodings have been observed~\cite{amplitude4,bush2016,Carbo15022017}.  Analyzing the combined use of the two codification strategies in this system, even within the framework of our simple model, would require the quantification of several parameters and this goes beyond the scope of
the present paper.  Yet, there is room for an insightful analysis, as the one that we follow in this Section. Namely, we keep the parameters of the transcription step in the values that we have used so far because they allow a good information transmission for frequency encoding (within a frequency range that, as shown in what follows, overlaps with the one observed in the yeast mating response pathway) and that, by changing them, MI would not vary significantly for amplitude encoding~\cite{givre_sci_rep_2023}. Then, we focus on the parameters of the $I_{ext}$-TF mapping. Relating them to experimental observations, we study whether the range of distinguishable stimuli can be expanded through the combination of frequency and amplitude encoding. We present this analysis in what follows. 

The canonical response of haploid mating type {\bf a} {\it
  S. cerevisiae} cells to the pheromone ($\alpha$-factor) secreted by
their potential mating partners, involves amplitude encoding as in
Eq.~(\ref{ATF}) with $h\sim 1$ and dimensional $EC_{50}\approx
3-5nM$~\cite{amplitude4,ColmanLerner2005,bush2016}.  Let us then
consider that the curve with $h=1$ in Fig.~\ref{fig:MI_amp}~(b) (asterisks)
represents this situation. Given that for this curve the dimensionless
$EC_{50}$ is $0.5$, we need to introduce the transformation
[$\alpha$-factor]= 6-10$nM\, I_{ext}$ to make the equivalence between
our results and the experiments.
 We then conclude from Fig.~\ref{fig:MI_amp}~(b) that MI decreases by $\sim 1$ bit ($\sim 40$\%) as the median
 [$\alpha$-factor] increases from 1.6-2.6 to 4-6.8$nM$ and by $\sim 1.5$ bit ($\sim 60$\%) if it increases up to 4.8-8$nM$. 

These cells also display intracellular \ca\ pulses of increasing
frequency with increasing pheromone concentration.  \ca\ pulses occur
very rarely for [$\alpha$-factor]=0 and their mean frequency increases
with [$\alpha$-factor] to values that become indistinguishable for
[$\alpha$-factor]$>10nM$~\cite{Carbo15022017}. Although the role of
these pulses in the pheromone response pathway is not clear yet, it is
conceivable that the nuclear localization of some of the TFs involved
in the response be pulsatile as well as it has been observed in the
response to \ca\ stress in yeast which involves intracellular
\ca\ pulses and the pulsatile nuclear localization of the TF,
Crz1~\cite{Cai2008}. A rough estimate in the form of Eq.~(\ref{FTF})
derived from Fig.~3C of~\cite{Carbo15022017} gives $T_{min}\approx
(10-12)$min, $\kappa\approx 8-9$ and a dimensional $b\sim
(0.4-0.5)/nM$. Let us consider that the curve plotted with asterisks
in Fig.~\ref{fig:MI_freq}~(a) ($T_{min}=10min$, $\kappa=9$, $b=4$)
corresponds to this situation. Given that the dimensionless $b$ for
this curve is 4 we need to introduce the transformation
[$\alpha$-factor]= (8-10)$nM\, I_{ext}$ to make the equivalence
between our results and the experiments. We then conclude from Fig.~\ref{fig:MI_freq}~(a) that MI
differs from its maximum by less than 20\% for the whole support of
the [$\alpha$-factor] distribution, $[0,8-10nM]$.

Under physiological conditions, distinguishing relatively subtle
differences in [$\alpha$-factor] is important for the cell to grow
towards the largest [$\alpha$-factor] regions to encounter its
potential partner.  If the partners are apart from one another, we can
expect that amplitude encoding be used at the earliest stages of the
detection, on one hand, because, as illustrated by the curve with
$h=1$ of Fig.~\ref{fig:MI_amp}~(b), it works correctly for relatively
small median concentrations.  On the other hand, because if [$\alpha$-factor]
is too low it will take a relatively long time for an individual cell
to collect enough statistics and ``respond'' correctly using frequency
encoding (our estimate of the mean interpulse time derived
from~\cite{Carbo15022017} yields $\sim 70min$ at
[$\alpha$-factor]=$1nM$). Furthermore, there is a time lag between
exposing the cells to $\alpha$-factor and the occurrence of
\ca\ pulses which, at the saturating level [$\alpha$-factor]=$100nM$,
is of 30min on average~\cite{Carbo15022017}.  As cells change their
form, getting closer to their partners, the pheromone concentration
around the growing mating projection gets larger. Amplitude encoding
might then cease to discriminate [$\alpha$-factor] values, but frequency
encoding could still do its job. Therefore, the apparently redundant use
of amplitude and frequency encoding to mount the pheromone response might
serve the purpose of allowing the cell to detect differences in [$\alpha$-factor] across different concentration ranges.

\section*{Summary, discussion and conclusions}
\label{sec:conclusions}

In this paper we compared two strategies commonly used by cells to
encode changes in the environment and generate responses: amplitude
and frequency encoding. While in the former increasing stimulus
intensities are transduced into increasing concentrations or
activation levels of the intermediaries of the pathway, in the latter,
pulsatile behaviors are induced in which the frequency increases with
the stimulus strength.  We had previously studied the information
capabilities of the transcription step when the Transcription Factor's
(TF) nuclear fraction displayed one or the other dynamics. We found that
the main difference between the two strategies was due to their
different sensitivity to changes in the promoters
properties~\cite{givre_sci_rep_2023}. In the present paper we focused
on the effect that the transduction of the external stimulus into the
TF's nuclear fraction had on the information transmission capabilities
for each codification strategy. To this end we assumed that amplitude
encoding entailed a mapping from the external stimulus, $I_{ext}$, to
the TFs mean concentration, $A_{TF}$, in the form of a Hill curve
(Eq.~(\ref{ATF})), an expression that describes many gene input
functions~\cite{alon_book} including those involved in the canonical pheromone
 response pathway in
yeast~\cite{amplitude4,ColmanLerner2005,bush2016}. For frequency
encoding we assumed that the interpulse time, $\tau$, was the sum of a
constant, $T_{min}$, and an exponentially distributed variable,
$\eta$, (Eq.~(\ref{tau_random})) with a mean, $T_{IP}$, that depended
exponentially on $I_{ext}$ (Eq.~(\ref{FTF})) as observed
experimentally in sequences of intracellular
\ca\ pulses~\cite{Thurley2014} and derived theoretically for different
classes of noisy driven excitable
systems~\cite{Eguiamindlin2000,Givre2024}. Screening a set of
$I_{ext}$ distributions defined over the same compact support, of
similar variance but different medians, we studied how the mutual
information, MI, between the mRNA produced over a finite time, $O$
(Eq.~(\ref{eq:Out})), and the stimulus strength, $I_{ext}$, varied with
the median of the distribution, Med$(I_{ext})$. We performed this
analysis under the assumption that Med$(I_{ext})$ was representative
of the stimulus strengths that constituted the bulk of each $I_{ext}$
distribution.

For amplitude encoding we found that the maximum MI is achieved if
Med$(I_{ext})$ approximately matches the Hill function's $EC_{50}$ in
those cases with cooperativity index, $h>2$,
(Fig.~\ref{fig:MI_amp}~(a)) while it is optimal at small values of
Med$(I_{ext})$ if $h=1$ (Fig.~\ref{fig:MI_amp}~(b)). Although
Eqs.~(\ref{A_random})--(\ref{ATF}) define only the first step in the
generation of the output (Eq.~(\ref{eq:Out})), we see that the
properties of the Hill function imprint their mark on the final
depedendence of MI with Med$(I_{ext})$.  Namely, the values
Med$(I_{ext})$ that give maximum MI roughly correspond to those of
$I_{ext}$ that yield maximum variability of the Hill function which,
in turn, satisfy Eq.~(\ref{r_A}), the condition under which two
stimuli that differ by $\Delta_I$ lead to distinguishable nuclear TF
concentrations.  We found a similar situation in the case of frequency
encoding in the sense that the properties of the first step of the
response generation (Eqs.~(\ref{tau_random})--(\ref{FTF})) highly
influenced the dependence of MI with Med$(I_{ext})$. In this case two
stimulus strengths that differ by $\Delta_I$ lead to distinguishable
interpulse frequency distributions if Eq.~(\ref{r_F}) is satisfied for
some $p>1/2$. $\Delta_F$ in Eq.~(\ref{r_F}) and the KL divergence (Eq.~(\ref{eq:KL})) are scale invariant in the
sense that they do not depend on $I_{ext}$, but only on
$\Delta_I$. This is the reason that underlies the weak dependence of
MI with Med$(I_{ext})$ of Fig.~\ref{fig:MI_freq}~(a). The timescales
of the subsequent steps in the generation of the response put limits
to this scale invariance.  As illustrated in
Fig.~\ref{fig:MI_freq}~(b), MI can decrease for decreasing
Med$(I_{ext})$ if the probability of eliciting one TF pulse during the
observation time, $T$, becomes to small. MI can also decrease for
increasing Med$(I_{ext})$. The analysis of the Med$(I_{ext})$ values
that yield maximum MI in each of the examples of
Fig.~\ref{fig:MI_freq}~(a) led us to conclude that the limiting
timescale in the high frequency end is that of mRNA degradation. This
coincides with our previous studies which showed that this timescale
limits the information transmitted through the transcription step
(Fig. 3D of~\cite{givre_sci_rep_2023}).

The above discussion shows that the qualitative difference between
amplitude and frequency encoding derives from the qualitative
differences between the Hill and the exponential functions that we
used to model the first step in the generation of the response. The
choice of the Hill function is incontestable. In the case of the exponential,
we  provided experimental evidence~\cite{Thurley2014,Cai2008} 
and cited theoretical works that derive this
dependence for different types of noise-driven excitable
systems~\cite{Eguiamindlin2000, Givre2024}. Given the widespread
presence of excitable dynamics in biology, including the paradigmatic
example in which spikes are used to transmit information (neurons), we
can expect a pervasive presence of this dependence in pulse-signaling
systems. Another important feature of the
exponential dependence is that it can yield relatively large ranges of mean
interpulse times~\cite{Givre2024}, as large as those observed in
experiments~\cite{Cai2008,Thurley2014}.  This feature is particularly
advantageous for frequency encoding as noticed
in~\cite{SchusterStefanMarhl}.

In the paper we also discussed the pheromone response pathway {\it
  S. cerevisiae} as a potential example in which the combination of frequency
and amplitude encoding could enlarge the range of external stimuli
({\it i.e.}, pheromone concentrations) over which the cells could
reliably distinguish different values. In this case, \ca\ pulses were
observed in mating type {\bf a} (MATa) cells in the presence of the
pheromone, $\alpha$-factor. Under physiological conditions, these
cells secrete the pheromone {\bf a}-factor which attracts mating type
$\alpha$ (MAT$\alpha$) cells. These two pheromones differ in various
properties, among them, their diffusivity, secretion mechanisms and
extracellular metabolism, so that their gradients around the secreting
cell can be expected to differ as well~\cite{biom12040502}. This led
to the hypothesis that the two pheromones conveyed different spatial
information to their potential partners, hypothesis that was
contradicted by recent observations~\cite{biom12040502}. Furthermore,
experiments in which MATa cells were exposed to different
$\alpha$-factor gradients showed that they could be decoded for a wide
range of mean $\alpha$-factor concentrations~\cite{moore_2008,
  dyer_2013}.  The ability of MATa cells to detect gradients, {\it
  i.e.}, to distinguish [$\alpha$-factor] values, across mean
concentrations was further confirmed by experiments in which the
gradients were produced with different source
strengths~\cite{jacobs_2022}. As concluded in~\cite{biom12040502}, it
seems that the cells ``do not rely on a narrow concentration range of
pheromone''.  As discussed in this paper, frequency encoding can endow
the cells with such scale-invariant discrimination ability. We expect
to do experiments to analyze this hypothesis in the future.

\bibliography{frontiers_alan}

\section*{Author contributions}
AG and SPD performed numerical and analytical calculations, respectively; AC-L and SPD conceived project; AC-L and SPD wrote paper

\section*{Acknowledgments}
This research has been supported by
UBA (UBACyT 20020170100482BA) and ANPCyT (PICT 2018-02026 and PICT-2021-III-A-00091 to SPD and
PICT 2019-1455 to ACL). 

\section*{Competing Interests} All authors declare no financial or non-financial competing interests.

\section*{Code availability}
The underlying code for this study is available in Git-Hub and can be accessed via this link: \url{https://github.com/alangivre/qualitatively-scripts}.

\end{document}